\documentclass{article}
\usepackage{spconf,amsmath,graphicx, siunitx, graphicx,subcaption,multicol, tabularx, multirow, array}
\usepackage{adjustbox}

\title{THE MIRRORNET : LEARNING AUDIO SYNTHESIZER CONTROLS INSPIRED BY SENSORIMOTOR INTERACTION}
\name{Yashish M. Siriwardena$^1$, Guilhem Marion$^2$ , Shihab Shamma$^{1, 2}$}

\address{$^{1}$ Institute for Systems Research, University of Maryland College Park, USA \\ $^{2}$Laboratoire des Systèmes Perceptifs, École Normale Supérieure, PSL University, France}

\begin{document}
\ninept
\vspace{-5pt}
\maketitle
\vspace{-5pt}
\begin{abstract}
Experiments to understand the sensorimotor neural interactions in the human cortical speech system support the existence of a bidirectional flow of interactions between the auditory and motor regions. Their key function is to enable the brain to `learn' how to control the vocal tract for speech production. This idea is the impetus for the recently proposed “MirrorNet”, a constrained autoencoder architecture. In this paper, the MirrorNet is applied to learn, in an unsupervised manner, the controls of a specific audio synthesizer (DIVA) to produce melodies only from their auditory spectrograms. The results demonstrate how the MirrorNet discovers the synthesizer parameters to generate the melodies that closely resemble the original and those of unseen melodies, and even determine the best set parameters to approximate renditions of complex piano melodies generated by a different synthesizer. This generalizability of the MirrorNet illustrates its potential to discover from sensory data the controls of arbitrary motor-plants.

\end{abstract}
\vspace*{-8pt}
\section{Introduction}
\label{sec:intro}
\vspace{-5pt}

Most organisms function by coordinating and integrating sensory signals with motor actions to survive and accomplish their desired tasks. For instance, visual and auditory signals guide animals to navigate their surroundings \cite{Wolpert2000ComputationalPO, KELLER2012809}. Similarly, auditory and proprioceptive percepts are essential in skilled tasks like playing the piano or speaking. The difficulty of learning to perform these tasks is enormous. It stems from the fact that to control such actions, one needs harmoniously to close the loop between sensing and action. That is, it is necessary to map the desired sensory signals to the correct commands, which in turn produce exactly the desired sensory signals when executed. 


\let\thefootnote\relax\footnote{© 2021 IEEE. Personal use of this material is permitted. Permission from
IEEE must be obtained for all other uses, in any current or future media,
including reprinting/republishing this material for advertising or promotional
purposes, creating new collective works, for resale or redistribution to servers
or lists, or reuse of any copyrighted component of this work in other works.}

But to learn the necessary mappings and interactions between the perception and action domains, standard Artificial Intelligence (AI) methodology typically relies on creating large databases that map the input sensory data to their corresponding actions, and then train intervening Deep Neural Networks (DNN) to associate the two domains \cite{machines_DNN, imitation_RL_for_robots}. 
Humans and animals however never learn complex tasks in this way. For instance, human infants learn to speak by first going through a “babbling” stage as they learn the “feel” or the range and limitations of their articulatory commands. They also listen carefully to the speech around them, initially implicitly learning it without necessarily producing any of it. When infants are ready to learn to speak, they utter incomplete malformed replica of the speech they hear. They also sense these errors (unsupervised) or are told about them (supervised) and proceed to adapt the articulatory commands to minimize the errors and slowly converge on the desired auditory signal. In other words, learning these complex sensorimotor mappings proceeds simultaneously and often in an unsupervised manner by listening and speaking all at once \cite{mirrorNetpaper, bird_paper, Kuhl2004EarlyLA}.


 Motivated by such learning of complex sensorimotor tasks, a new autoencoder architecture, referred to as the “Mirror Network” (or MirrorNet) was recently proposed in Shamma et al. \cite{mirrorNetpaper}. The essence of this biologically motivated  algorithm is the bidirectional flow of interactions (`forward' and `inverse' mappings) between the auditory and motor responsive regions, coupled to the constraints imposed simultaneously by the actual motor plant to be controlled. In this paper we extend and demonstrate the efficacy of the MirrorNet architecture in learning audio synthesizer controls/parameters to synthesize a melody of notes using a commercial, widely available synthesizer (DIVA) developed by U-He\footnote[1]{https://u-he.com/products/diva/}. 

MirrorNet is fundamentally different from the Differentiable Digital Signal Processing (DDSP) based models \cite{ddsp_original, ddsp_paper2} which effectively learn a differentiable music synthesizer, whereas the goal of the MirrorNet is to learn controls to drive a given non-differentiable, off-the-shelf music synthesizer. Previous work with DNNs on determining music and speech synthesizer controls are all based on at least partially supervised techniques which often involve large databases of audio and control parameter pairs (order of 1000s) \cite{music_synthesizer_paper2, music_synthesizer_IRCAM, VAE_Dexed_FM, georges21_interspeech_vowels}. Furthermore, previous efforts have mostly demonstrated the ability to compute the controls for single notes or single vowels for speech \cite{music_synthesizer_IRCAM, saha_vowel_synthesis}. In this paper we propose an alternative approach model which is fundamentally unsupervised, in that it does not require matched pairs of input melodies and their corresponding control parameters. The proposed model can predict synthesizer controls for a melody composed of several notes demonstrating the scalability of the model for real world applications. The true potential of the MirrorNet is further validated by showing how well it can predict synthesizer controls not only for DIVA generated melodies, but for other off-the-shelf synthesizer-generated melodies.

\vspace*{-8pt}
\section{MirrorNet Model}
\label{sec:ModelArchitecture}
\vspace*{-5pt}
\subsection{Model Architecture}
\vspace*{-2pt}
\label{ssec:model_archi}

\begin{figure}[t]
    \centering
    \begin{subfigure}{0.9\columnwidth}
        \includegraphics[width=\linewidth, scale=0.9, height=20mm]{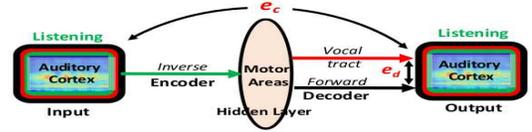}
        \subcaption{MirrorNet: Autoencoder Architecture}
        \label{fig:model_archi}
    \end{subfigure}
    \begin{subfigure}{0.9\columnwidth}
        \includegraphics[width=\linewidth, scale=0.9, height=30mm ]{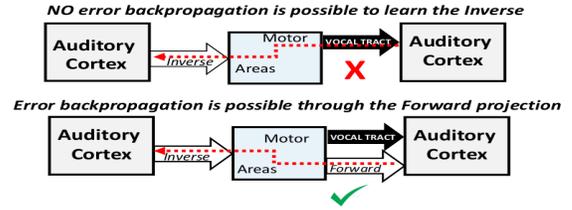}
        \subcaption{Role of the forward pass}
        \label{fig:model_backprop}
    \end{subfigure}
    \caption{MirrorNet Model Architecture for speech and the critical role of the forward projection (taken from \textit{Learning Speech Production and Perception through Sensorimotor Interaction} by Shamma et al. in \textit{Cerebral Cortex Communications.})}
    \label{fig:Model_architecture}
\end{figure}

The  MirrorNet was initially proposed as a model for learning to control the vocal tract and is based on an autoencoder architecture. The structure of this network is shown in Figure \ref{fig:model_archi} \cite{mirrorNetpaper}, depicting the biological structures and experiments that motivated the network. The goal of the model is to learn two neural projections, an inverse mapping from auditory representation to motor parameters (Encoder) and a forward mapping from the motor parameters to the auditory representation (Decoder). For simplicity, we use auditory spectrograms \cite{Wang_auditory_spec} generated from the audio streams as the input and output representations, but other representations may prove more versatile (e.g., cortical representations \cite{Chi_cortical_rep}). The “motor” parameters in this study are the parameters needed to synthesize the closest possible audio signals matching the inputs. The primary difference between this MirrorNet and the previously studied model in \cite{mirrorNetpaper} is the use of the music synthesizer (DIVA) with its unique set of parameters. 

As shown in Figure \ref{fig:model_archi}, the MirrorNet model is optimized simultaneously with two loss functions namely the `encoder loss'($e_{c}$) and the `decoder loss'($e_{d}$). The encoder loss is the typical autoencoder loss - the Mean Squared Error (MSE) between the input auditory spectrogram and the reconstructed auditory spectrogram from the decoder (forward path). The decoder loss is the MSE between the auditory spectrograms generated by the DIVA (the motor plant path) and the decoder (forward path). It is the `decoder loss' that constrains the latent space to converge to the expected control parameters while simultaneously reducing ($e_{c}$), and this is the key feature of the MirrorNet architecture.

Figure \ref{fig:model_backprop} shows the role of the `forward' path in the model, namely to back-propagate the errors computed to learn the `inverse' mapping and hence the control parameters. In general, directly computing a vocal-tract or an audio synthesizer inverse is difficult if not impossible because of its complexity, nonlinearity, and our incomplete knowledge of its workings. The MirrorNet in Figure \ref{fig:model_backprop} (bottom panel) solves this problem by adding the forward projection that serves as a parallel, “neural” model of the vocal tract or the audio synthesizer, or any motor-plant to be used. The critical importance of this “neural” projection is that it readily provides a route for the $e_{c}$ errors to back-propagate to the motor areas (latent space), enabling the training of the inverse mapping (Encoder).

\vspace*{-10pt}
\subsection{Model Implementation and Training}
\label{ssec:model_impl}
\vspace*{-2pt}

\begin{figure*}[t]
  \centering
  \includegraphics[width=1.0\linewidth,height=50mm]{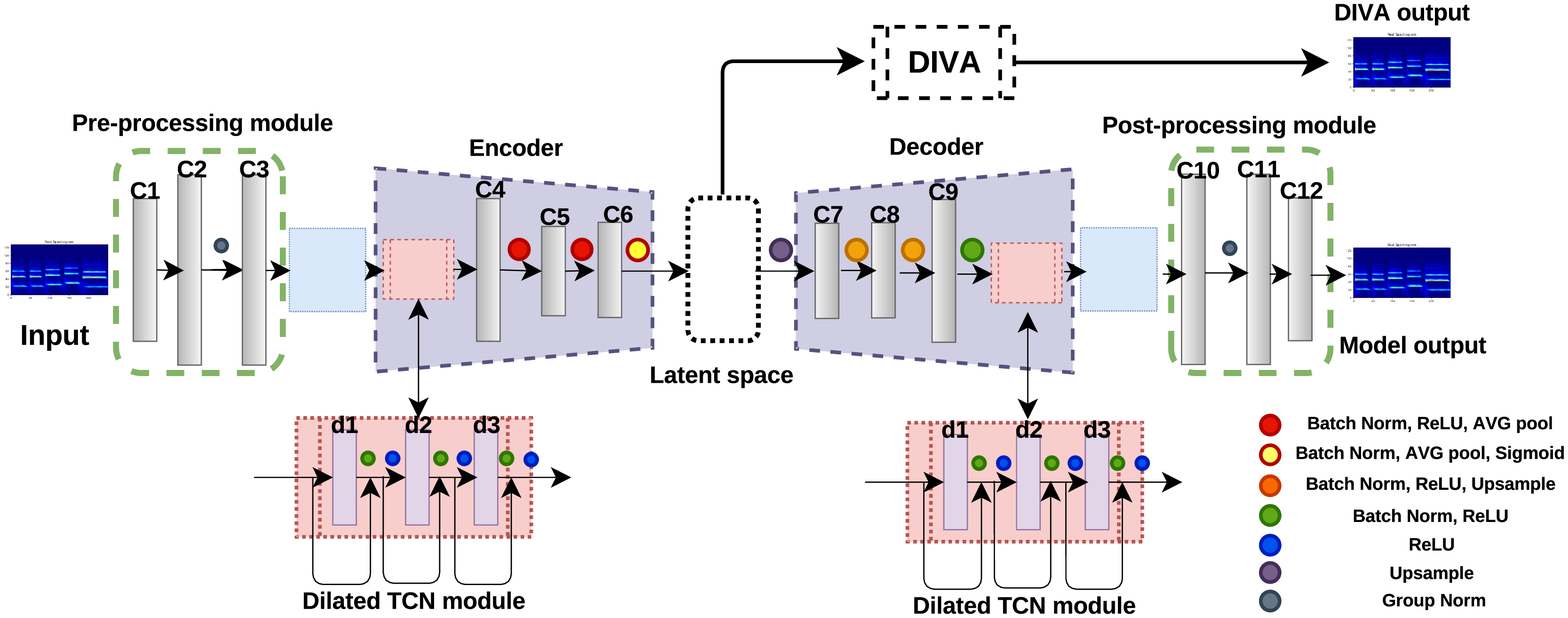}
  \caption{DNN architecture of the MirrorNet model. Here C1-C12 represent 1D-CNN layers where d1-d3 represent 1D dilated CNN layers.}
    \label{fig:DNN_model}
\end{figure*}

The MirrorNet for audio synthesizer control is implemented in PyTorch with 1-D convolutional (CNN) layers modeling both the encoder and decoder. The complete network is inspired by the multilayered Temporal Convolution Network (TCN) \cite{Lea_2017_Temporal_CNN}. Figure \ref{fig:DNN_model} shows the complete DNN model architecture with its sub-modules used for pre/post processing and dilated TCN. The pre/post processing modules are symmetrically matched (C1$\equiv$C12, C2$\equiv$C11, C3$\equiv$C10) and have 128, 256 and 256 filters respectively with 1$\times$1 kernels. d1, d2 and d3 dilated CNN layers have a kernel size of 3 with 1,4 and 16 dilation rates respectively. The CNN layers in the encoder and decoder are also symmetrically matched and the C4, C5 and C6 layers have 256, 128 and 7 filters respectively with 1$\times$1 kernels. The latent space dimensions are chosen to match with the number of parameters to be learned and the number of notes in each melodic segment. For example to learn 7 controls of the DIVA synthesizer to generate a melodic segment of 5 notes, we use a latent space of (7$\times$5) dimensions. Average pooling is done after C4, C5 and C6 layers (window sizes of 5, 5 and 2 respectively) while upsampling is done before C7, C8 and C9 layers (window size of 2, 5 and 5 respectively). The auditory spectrograms used as inputs (and outputs) of the model are of dimension (128$\times$250). We use auditory spectrograms which have a logarithmic frequency scale, simply because they provide a unified multi-resolution representation of the spectral and temporal features likely critical in the perception of sound \cite{Wang_auditory_spec, Chi_cortical_rep}.

Unlike a regular autoencoder, the MirrorNet is trained in two alternating stages in each iteration. The decoder is trained first (to minimize $e_{d}$) for a chosen number of epochs. Then, the encoder is trained (to minimize $e_{c}$) for a given number of epochs and this alternation of training is continued until both  losses converge to a minimum. Learning rates of 1e-2 and 1e-3 were used for the encoder and decoder networks, respectively. The best learning rates were determined based on a grid search testing all the combinations from [1e-2, 1e-3, 1e-4, 3e-4] for both the encoder and decoder which result in the lowest training errors at convergence. The two objective functions were optimized using the ADAM optimizer with an `ExponentialLR' learning rate scheduler and a decay (gamma) of 0.5. All the models were trained using NVIDIA Quadro P6000 GPUs and on average the models converged after around 32 hours of training. For further implementation information of the network, the PyTorch project is publicly available in GitHub \footnote[2]{https://github.com/Yashish92/MirrorNet-for-Audio-synthesizer-controls}. Sample audio reconstructions can also be found in the supporting web page hosted in the GitHub repository.   


\vspace*{-10pt}
\subsection{DIVA audio synthesizer}
\vspace*{-2pt}
\label{ssec:Diva}

We use DIVA, an off-the-shelf commercial synthesizer as our audio synthesizer for the MirrorNet model. DIVA has almost all its parameters MIDI-controlled. A python library called RenderMan \footnote[3]{https://github.com/fedden/RenderMan} is used to batch-generate audio files using a fixed set of parameters. We built a software layer with RenderMan to drive DIVA to synthesize a melody of notes by concatenating individual notes synthesized by DIVA. All the melodies used in this paper are 2 seconds long and sampled at 44.1 kHz. The parameters are all continuous and normalized between [0,1]. Table \ref{tab:diva_param} lists the set of parameters selected for the learning experiments with the MirrorNet, and the corresponding parameter labels from DIVA where applicable.

\vspace*{-2pt}
\begin{table}
\centering
\normalsize
\caption{Set of Audio controls/parameters used. Here MIDI note and MIDI duration are parameters set in RenderMan library to drive the synthesizer patch.}
\vspace{-5pt}
\label{tab:diva_param}
\begin{tabular}{|l|l|}
\hline
\textbf{Parameter Name}     & \textbf{DIVA preset}     \\ \hline
MIDI note (Pitch)                  & -       \\ \hline
MIDI duration  & - \\ \hline
Volume                &OSC : Volume2   \\ \hline
Band pass filter (center frequency)                &VCF1: Frequency   \\ \hline
Filter Resonance                &VCF1: Resonance   \\ \hline
Envelope Attack               &ENV1: Attack   \\ \hline
Envelope Decay                &ENV1: Decay   \\ \hline
Vibrato Rate                  &LFO1: Rate \\ \hline
Vibrato Intensity             &OSC : Vibrato\\ \hline
Vibrato Phase                 &LFO1: Phase \\ \hline
\end{tabular}
\end{table}
\vspace{-5pt}

    
  

\vspace*{-2pt}
\section{Experiments and Results}
\label{sec:exps}
\vspace*{-4pt}

  
\subsection{Learning DIVA parameters from melodies synthesized with the same set of parameters (set 1) }
\vspace*{-2pt}
\label{ssec:exp1}

In this first experiment, we use 400 melodies (set 1) to train the MirrorNet and test with 80 melodies, all originally synthesized by DIVA. The advantage of this set of melodies is that we have its ground-truth parameter values, and hence we can assess how accurately the MirrorNet rediscovers them and reconstructs the melodies. Each melody contains 5 notes and is 2 seconds long. The train and test set of melodies were synthesized by randomly sampling a total of 7 parameters (the first 7 parameters in Table \ref{tab:diva_param}) using a defined range and keeping a pre-defined set of other parameters constant across all notes and melodies. The pre-defined set of parameters used for the experiments can be found in the GitHub repository of the project.

\begin{figure}[t]
  \centering
  \includegraphics[width=1.0\columnwidth]{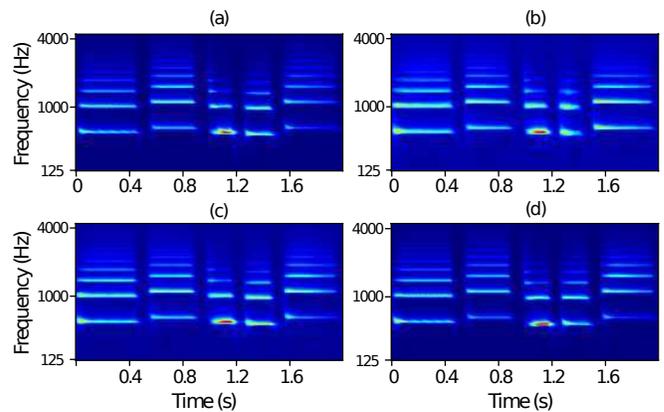}
  \caption{Auditory spectrograms from the model learned with DIVA synthesized melodies (set 1). (a) Input melody (b) Decoder output from true DIVA parameters (c) Final output from the decoder (d) DIVA output from the learned control parameters}
    \label{fig:spec_ex1}
\end{figure}

Figure \ref{fig:spec_ex1} depicts auditory spectrograms of a given melody at various stages in the fully-trained MirrorNet. The spectrogram (b) suggests how well the decoder has learned to generate an identical spectrogram to the one generated with DIVA for the exact same controls. The spectrogram (d) suggests how well predicted DIVA controls are from the encoder to synthesize an identical melody to the input.

\begin{figure}[t]
  \centering
  \includegraphics[width=1.0\columnwidth]{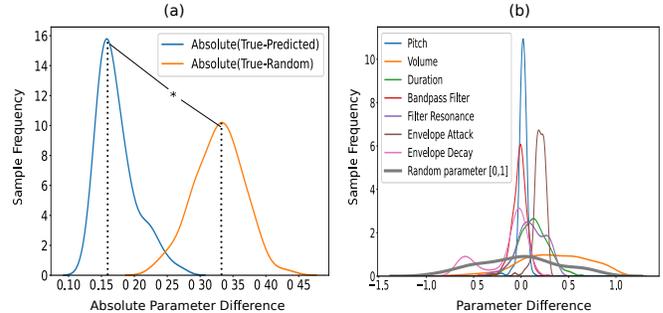}
  \caption{Evaluating statistical significance of the predicted DIVA parameters with respect to a set of random parameters on the test set (a) Distributions for absolute parameter differences across all parameters (b) Distributions of parameter differences (ground truth - predicted) for 7 parameters and the distribution for a random parameter difference (ground truth - random)}
    \label{fig:stat_fig}
\end{figure}

We performed preliminary statistical tests to evaluate the robustness of the MirrorNet in predicting the 7 parameters. Plot in Figure \ref{fig:stat_fig}a validates that the predicted and ground truth parameters are significantly closer together than would result from a random set of values. A second test was performed to check how well the predictions of each parameter are compared to a random prediction. For that we performed a Levene's test that confirmed that all parameter predictions were significantly better than chance. Plot in Figure \ref{fig:stat_fig}b shows the parameter difference distributions for the test set. The distributions also suggest that critical parameters like pitch, bandpass filter, filter resonance and duration are predicted with significant accuracy where as volume and envelope attack parameters are predicted with comparatively lower accuracy.


\vspace*{-10pt}

\subsection{Learning DIVA parameters from melodies synthesized with extra unknown DIVA parameters (set 2) }
\vspace*{-2pt}

\begin{figure}[t]
  \centering
  \includegraphics[width=1.0\columnwidth]{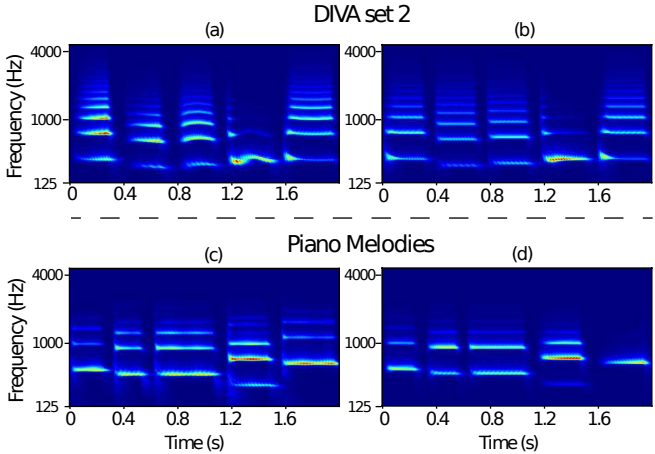}
  \caption{(Top panel) Auditory spectrograms from the model learned with DIVA synthesized melodies (set 2) (a) Input melody (b) DIVA output from the learned control parameters.
  (Bottom panel) Auditory spectrograms from the model learned with piano melodies. (c) Input melody (d) DIVA output from the learned control parameters.}
    \label{fig:spec_ex2_3}
\end{figure}

In this experiment, we use a train set of 400 and a test set of 80, both DIVA generated melodies (set 2) which are synthesized in similar fashion to set 1 except for the fact that they now use all the 10 parameters in Table \ref{tab:diva_param}. The MirrorNet is still trained to predict 7 parameters as in previous experiment. The goal here is to demonstrate that the MirrorNet can approximate the input melodies even if they have additional sound/musical qualities that are impossible for the restricted set of 7 DIVA parameters to reproduce, e.g., vibrato in this case. The top panel in Figure \ref{fig:spec_ex2_3} illustrates the original (vibrato) notes and the successfully regenerated melody captured with only 7 parameters (vibrato not included).  

\vspace*{-2pt}
\vspace*{-5pt}

\begin{table*}[t]
    \centering
    \normalsize
    \caption{Mean and variance of Mean Square Errors (MSE's) across multiple model training runs}
    \vspace{-8pt}
    \label{tab:MSE_table}
    \begin{tabular}{|l|l|l|l|}
    \hline
    \textbf{Input melody type}     & \textbf{Train/Test for Input vs DIVA(learned)} &\textbf{Parameter-Train} &\textbf{Parameter-Test}\\ \hline
    DIVA melodies (set 1)  &2.995$\pm$.21/3.596$\pm$.15       &0.0666$\pm$.003   &0.0671$\pm$.002 \\ \hline
    DIVA melodies (set 2)  &6.380$\pm$.34/8.101$\pm$.20     &0.0832$\pm$.007   &0.0857$\pm$.004 \\ \hline
    Piano melodies         &4.585$\pm$.25/4.751$\pm$.22      &-   & - \\ \hline
    \end{tabular}
\end{table*}

\vspace*{-5pt}
\subsection{Learning DIVA parameters to synthesize melodies generated from other synthesizers}
\vspace*{-2pt}
\label{ssec:exp3}

A fundamental advantage of the MirrorNet is its ability to discover the DIVA parameters corresponding to music generated by other sources and synthesizers by finding parameters that allow the DIVA output to be as close as possible, given the constraints of the number of parameters (here 7 are used), to the original input. The experiment utilized 400 5-notes long piano melodies of 2 seconds that are synthesized by a Fender Rhodes digital imitation (Neo-Soul Keys generated trough Kontakt 5). The network successfully reproduces accurate renditions of the piano music from unseen samples (test set of 80 samples) using the decoder/encoder mappings learned during the training. The bottom panel in Figure \ref{fig:spec_ex2_3} shows such an example where the DIVA produces a melody which closely resembles the input piano melody.




\vspace{-12pt}
\section{Discussion}
\label{sec:discus}
\vspace{-5pt}



We described a MirrorNet model inspired by cortical sensorimotor interactions measured when humans speak or play a musical instrument \cite{mirrorNetpaper}. The first two experiments utilized DIVA generated melodies for training, and this allowed us evaluate the effectiveness of the MirrorNet given the ground truth parameters to compare against, e.g., to perform preliminary tests to validate the MirrorNet predictions of the synthesizer controls across all the training and test sets, as shown in Table \ref{tab:MSE_table}. The MSE values for the test set compared to the train set in Table \ref{tab:MSE_table} also give an idea on how well the model generalizes for the unseen input melodies. 

Taking the MirrorNet to the next level in the last experiment, we demonstrated how the MirrorNet could closely approximate a set of controls for DIVA to synthesize a set of piano melodies generated by a completely different synthesizer. This idea opens up a whole new area of applications in music synthesis as it describes a tool to find parameters for an arbitrary synthesizer that maximally approximate an arbitrary sound without being necessarily capable to exactly reproduce it (reproduce a violin using a guitar for instance). It should also be noted that this paper only discusses results in synthesizing fixed duration melodies with a fixed number of notes, but it is a step in the right direction to synthesizing a piece of music which can have a variable number of notes in a fixed frame of audio.

The inspiration of the MirrorNet also comes from the area of computational neuroscience and especially to learning and predictive processing. Our brain is able to extract strong relations between sensory stimuli and their corresponding motor parameters that enable children to learn to speak by mere passive exposure to speech without any proper external teaching. In addition, after learning to control their own vocal tract, adults can, without any additional training, produce sounds they hear even if the acoustic target is not reachable by their specific vocal tract (case of the experiments 2 and 3). However, the brain is able to find a set of motor parameters that approximate well the target sound while being produced by the specific vocal tract. Such predictive mechanism can also be seen in music production when humans learn how to play an instrument by mapping the auditory stimulation to the motor commands to a specific instrument. Even music perception rely on similar predictive pathways where high-order cortical areas constantly predict activation in the auditory cortices in order to modulate attention and emotions, for instance\cite{Marion2021, DiLiberto}. 

Finally, from an engineering perspective, the MirrorNet can solve problems where it is hard to find a reasonable number of examples to train a regular feed-forward DNN network, or to learn from examples that may not be exactly similar to the motor-plant outputs, e.g., learning to synthesize a melody from naturally played music. We moreover believe that the MirrorNet can be generalized to design algorithms that can control motor-plants such as self-driving vehicles given various sensory data.





\vspace*{-8pt}
\section{Conclusion and Future Work}
\label{sec:conclusion}
\vspace{-5pt}

This paper presents an autoencoder architecture inspired by sensorimotor interactions to discover and learn audio synthesizer controls. The work is novel in that the proposed MirrorNet can learn the necessary controls to produce a melody in a completely unsupervised way. It can also be potentially generalized to learn the controls for any motor-plant action from the sensory data associated with them. However, to realize all these potentials, many more advances are needed. For example, for the audio synthesizer controls explored here, it is necessary to scale up the current implementations to far more parameters that capture richer aspects of the sound (e.g., vibrato), to deploy more advanced and richer representations of the sound beyond the spectrograms, to devise more efficient and faster training paradigms, and finally to target the synthesis of continuous musical melodies which can have a variable number of notes. 

\vspace{-8pt}
\section{ACKNOWLEDGEMENTS}
\label{sec:typestyle}
\vspace{-5pt}

This work was supported by Advanced ERC Grant NEUME 787836 and Air Force Office of Scientific Research and National Science Foundation grants to S.A.S.; and FrontCog Grant ANR-17-EURE-0017, PSL Idex ANR-10-IDEX-0001-02, and a PhD scholarship from the Research Chair on Beauty Studies PSL–L’Oréal to G.M.



\bibliographystyle{IEEEbib}
\bibliography{mybib}

\begin{thebibliography}{10}

\bibitem{Wolpert2000ComputationalPO}
D.~Wolpert and Zoubin Ghahramani,
\newblock ``Computational principles of movement neuroscience,''
\newblock {\em Nature Neuroscience}, vol. 3 suppl. 1, pp. 1212--1217, 2000.

\bibitem{KELLER2012809}
Georg B. Keller, Tobias Bonhoeffer, and Mark Hübener,
\newblock ``Sensorimotor mismatch signals in primary visual cortex of the
  behaving mouse,''
\newblock {\em Neuron}, vol. 74, no. 5, pp. 809--815, 2012.

\bibitem{machines_DNN}
Yiwei Fu, Devesh~K. Jha, Zeyu Zhang, Zhenyuan Yuan, and Asok Ray,
\newblock ``Neural network-based learning from demonstration of an autonomous
  ground robot,''
\newblock {\em Machines}, vol. 7, no. 2, 2019.

\bibitem{imitation_RL_for_robots}
Lei Tai and Ming Liu,
\newblock ``Deep-learning in mobile robotics - from perception to control
  systems: {A} survey on why and why not,''
\newblock {\em CoRR}, vol. abs/1612.07139, 2016.

\bibitem{mirrorNetpaper}
Shihab Shamma, Prachi Patel, Shoutik Mukherjee, Guilhem Marion, Bahar
  Khalighinejad, Cong Han, Jose Herrero, Stephan Bickel, Ashesh Mehta, and Nima
  Mesgarani,
\newblock ``{Learning Speech Production and Perception through Sensorimotor
  Interactions},''
\newblock {\em Cerebral Cortex Communications}, vol. 2, no. 1, 2020.

\bibitem{bird_paper}
Silvia Pagliarini, Arthur Leblois, and Xavier Hinaut,
\newblock ``{Canary Vocal Sensorimotor Model with RNN Decoder and
  Low-dimensional GAN Generator},''
\newblock in {\em 2021 IEEE International Conference on Development and
  Learning (ICDL)}, 2021, pp. 1--8.

\bibitem{Kuhl2004EarlyLA}
Patricia~K. Kuhl,
\newblock ``Early language acquisition: cracking the speech code,''
\newblock {\em Nature Reviews Neuroscience}, vol. 5, pp. 831--843, 2004.

\bibitem{ddsp_original}
Jesse Engel, Lamtharn Hantrakul, Chenjie Gu, and Adam Roberts,
\newblock ``Ddsp: Differentiable digital signal processing,'' 2020.

\bibitem{ddsp_paper2}
Jesse Engel, Rigel Swavely, Lamtharn Hantrakul, Adam Roberts, and Curtis
  Hawthorne,
\newblock ``Self-supervised pitch detection by inverse audio synthesis,''
\newblock 2020.

\bibitem{music_synthesizer_paper2}
Matthew~John Yee-King, Leon Fedden, and Mark d'Inverno,
\newblock ``Automatic programming of vst sound synthesizers using deep networks
  and other techniques,''
\newblock {\em IEEE Transactions on Emerging Topics in Computational
  Intelligence}, vol. 2, no. 2, pp. 150--159, 2018.

\bibitem{music_synthesizer_IRCAM}
Philippe Esling, Naotake Masuda, Adrien Bardet, Romeo Despres, and Axel
  Chemla-Romeu-Santos,
\newblock ``Flow synthesizer: Universal audio synthesizer control with
  normalizing flows,''
\newblock {\em Applied Sciences}, vol. 10, no. 1, 2020.

\bibitem{VAE_Dexed_FM}
Gwendal Le~Vaillant, Thierry Dutoit, and Sébastien Dekeyser,
\newblock ``Improving synthesizer programming from variational autoencoders
  latent space,''
\newblock in {\em Proceedings of the 24th International Conference on Digital
  Audio Effects (DAFx20in21)}, Sept. 2021.

\bibitem{georges21_interspeech_vowels}
Marc-Antoine Georges, Laurent Girin, Jean-Luc Schwartz, and Thomas Hueber,
\newblock ``{Learning Robust Speech Representation with an
  Articulatory-Regularized Variational Autoencoder},''
\newblock in {\em Proceedings Interspeech 2021}, 2021, pp. 3345--3349.

\bibitem{saha_vowel_synthesis}
Pramit Saha and Sidney Fels,
\newblock ``{Learning Joint Articulatory-Acoustic Representations with
  Normalizing Flows},''
\newblock in {\em Proceedings Interspeech 2020}, 2020, pp. 3196--3200.

\bibitem{Wang_auditory_spec}
Kuansan Wang and S.~Shamma,
\newblock ``Self-normalization and noise-robustness in early auditory
  representations,''
\newblock {\em IEEE Transactions on Speech and Audio Processing}, vol. 2, no.
  3, pp. 421--435, 1994.

\bibitem{Chi_cortical_rep}
Taishih Chi, Powen Ru, and Shihab~A. Shamma,
\newblock ``Multiresolution spectrotemporal analysis of complex sounds,''
\newblock {\em The Journal of the Acoustical Society of America}, 2005.

\bibitem{Lea_2017_Temporal_CNN}
Colin Lea, Michael~D. Flynn, Rene Vidal, Austin Reiter, and Gregory~D. Hager,
\newblock ``Temporal convolutional networks for action segmentation and
  detection,''
\newblock in {\em Proceedings of the IEEE Conference on Computer Vision and
  Pattern Recognition (CVPR)}, July 2017.

\bibitem{Marion2021}
Guilhem Marion, Giovanni~M. Di~Liberto, and Shihab~A. Shamma,
\newblock ``The music of silence: Part i: Responses to musical imagery encode
  melodic expectations and acoustics,''
\newblock {\em Journal of Neuroscience}, vol. 41, no. 35, pp. 7435--7448, 2021.

\bibitem{DiLiberto}
Giovanni~M. Di~Liberto, Guilhem Marion, and Shihab~A. Shamma,
\newblock ``The music of silence: Part ii: Music listening induces imagery
  responses,''
\newblock {\em Journal of Neuroscience}, vol. 41, no. 35, pp. 7449--7460, 2021.

\end{thebibliography}

\end{document}